\documentclass[9pt,twocolumn,twoside]{opticajnl}
\journal{opticajournal} 

\setboolean{shortarticle}{true}

\usepackage{lineno}
\usepackage{physics}

\title{Generation of 480~nm picosecond pulses for ultrafast excitation of Rydberg atoms}

\author[1,2]{T.P. Mahesh}
\author[1]{Takuya Matsubara}
\author[1]{Yuki Torii Chew}
\author[1,2]{Takafumi Tomita}
\author[1,*]{Sylvain de Léséleuc}
\author[1,2]{Kenji Ohmori}

\affil[1]{Institute for Molecular Science, National Institutes of Natural Sciences, Okazaki 444-8585, Japan }
\affil[2]{SOKENDAI (The Graduate University for Advanced Studies), Okazaki 444-8585, Japan }

\affil[*]{sylvain@ims.ac.jp}

\begin{abstract}
Atoms in Rydberg states are an important building block for emerging quantum technologies. While the excitation to the Rydberg orbitals are typically achieved in more than tens of nanoseconds, the physical limit is in fact much faster, at the ten picoseconds level. Here, we tackle such ultrafast Rydberg excitation of a Rubidium atom by designing a dedicated pulsed laser system generating 480~nm pulses of 10~ps duration. In particular, we improved upon our previous design by using an injection-seeded optical parametric amplifier (OPA) to obtain stable pulsed energy, decreasing the fluctuation from 30~\% to 6~\%. We then succeeded in ultrafast excitation of Rydberg atoms with excitation probability of $\sim$90~\%, not limited anymore by energy fluctuation but rather by the atomic state preparation, addressable in future works. This achievement broadens the range of applications of Rydberg atoms.
\end{abstract}

\setboolean{displaycopyright}{false} 

\begin{document}

\maketitle

Rydberg atoms, with one of their valence electron in a loosely bound state, have giant orbitals with a radius of $\sim$100~nm. Consequently, they act as giant antenna with exaggerated sensitivity to electric fields, of great use in quantum science. 
They have been employed for sensing microwave~\cite{anderson2021self,sedlacek2012microwave} and THz radiation~\cite{Weatherill2020,nill2023avalanche}, even down to the quantum regime of a single microwave photon in a cavity~\cite{RevModPhys.85.1083,Kumar2023}.
Furthermore, these Rydberg atoms can sense each other through the dipole-dipole interaction, which is the basis of Rydberg-atom-based quantum optics~\cite{cantu2020repulsive}, quantum simulators~\cite{browaeys2020many}, and quantum computers \cite{graham2022multi,bluvstein2022quantum}.

One of the challenge with Rydberg atoms is the difficulty in exciting the electron from its ground state into a giant Rydberg orbit, as this requires coherent radiation in the visible to UV range with high intensity. 
Following two decades of development, continuous wave (cw) laser technology have now proven transfer probability of better than 99~\% to Rydberg states~\cite{evered2023high, Scholl2023, Ma2023}. Because of the current limits in cw laser power and switching speeds of modulators, the fastest excitation are performed in 10 to 100~ns. Accordingly, the atomic resonance needs to be within 10~MHz of the laser frequency. While this is easily achieved in the pristine conditions offered by most cold-atom experiments, this condition precludes excitation of thermal atoms (because of Doppler shift) or of atoms close to surfaces (because of stray electric field).   
Furthermore, when exciting more than one atom, Rydberg atoms perturb each other because of their strong interaction. This Rydberg blockade effect~\cite{browaeys2020many}, while immensely useful in a lot of applications, is also a limit when trying to prepare a dense ensemble of Rydberg atoms, or even just two Rydberg atoms at a micrometer distance~\cite{chew2022ultrafast}.

\begin{figure}[]
\centering
\includegraphics[width=\linewidth]{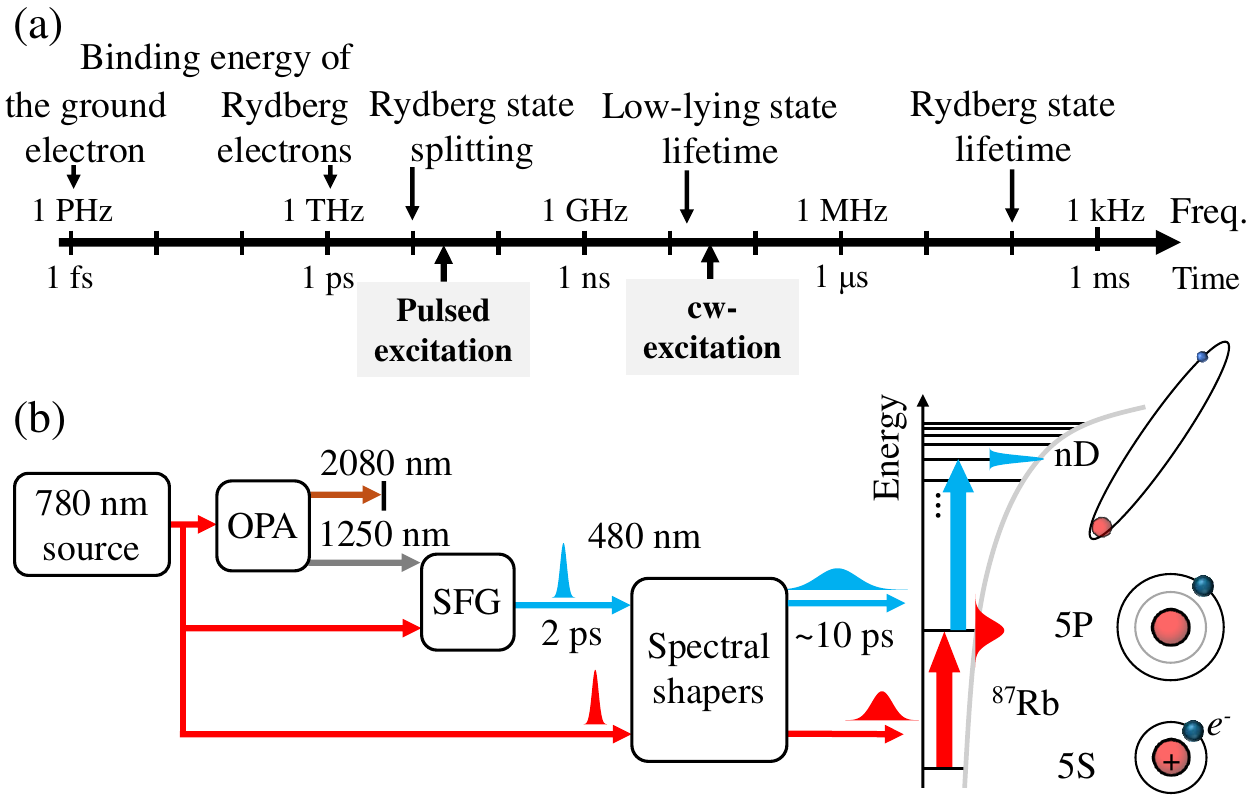}
\caption{(a) Rydberg timescale with some physical properties. (b, right) Electronic structure of a Rubidium atom, showing the valence electron in the $5S$ ground state, the first excited $5P$ level and the giant Rydberg $nD$ orbits. The schematics are not to scale, as the Rydberg states are 1000 times larger than the $5S$ orbit. (left) Pulsed laser system for ultrafast excitation. A 780~nm source emits a 2~ps pulse, an OPA and SFG convert it to 480~nm, and a spectral shaper reduces its bandwidth to 30~GHz for resolving the Rydberg states. The 780~nm and 480~nm pulses are shone one after the other.}
\label{fig:image1}
\end{figure}

Remarkably, it is physically allowed to excite Rydberg atoms much faster than discussed above. The speed limit is first set by the binding energy, $E_{n} = -13.6/n^2$~eV ($n$ is the principal quantum number), limiting excitation timescale to picosecond or longer to avoid photo-ionizing the electron. 
As shown in Fig.~\ref{fig:image1}(a), a second more restrictive limit is the splitting between neighbouring Rydberg states, 100~GHz around $n = 40$, setting the speed limit at 10~ps. To enter such ultrafast timescale, one has to turn from cw lasers to pulsed lasers. 
Nanosecond pulses, from Q-switched lasers, have for example enabled Rydberg excitation of thermal atoms (despite the Doppler broadening)~\cite{huber2011ghz}. For even shorter duration, one relies instead on mode-locked lasers delivering sub-picosecond pulses (the pulses can be later spectrally engineered to extend them to 10~ps).

Our research group have been following this approach, with a first demonstration of populating 10~\% of a single Rydberg state with a 10~ps pulse~\cite{takei2016direct}. This unlocked the investigation of Gigahertz interaction strength~\cite{takei2016direct,bharti2023picosecond,bharti2023strong}, as well as overlapping Rydberg orbitals~\cite{mizoguchi2020ultrafast}. In these first results, atoms are excited from the $5S$ ground state of $^{87}$Rb to a $nD$ Rydberg state by virtually going through the intermediate $5P$ state: the standard two-photon off-resonant scheme for Rydberg excitation. This scheme is justified when the excitation timescale is longer than the short lifetime of the $5P$ state (25~ns), which is the case for excitation with cw-lasers. Recognizing that, for sub-nanosecond excitation, the intermediate $5P$ state is \textit{metastable}, we can switch to a more power-efficient scheme where the atom is first transferred fully from $5S$ to $5P$ with a first 780~nm pulse, and then from $5P$ to $nD$ with a second 480~nm pulse. Applying this sequential two-photon excitation, we could observe Rabi oscillation with a greatly increased excitation probability of 75~\%~\cite{chew2022ultrafast}.
We also identified that the remaining error could be largely attributed to large pulse-to-pulse energy fluctuation of 30 $\%$ of the 480~nm pulse.
In this letter, we report the development of a new pulse laser system to further improve the Rydberg excitation fidelity.
The key upgrade is
the injection seeding of the optical parametric amplifiers~\cite{traub2011efficient,kohler20029,homann2013seeding,britton1998wavelength}. This largely decreased the pulse-to-pulse energy fluctuation and led us to achieve a maximum Rydberg excitation probability of $\sim 90$~\%.

\begin{figure}[]
\centering
\includegraphics[width=\linewidth]{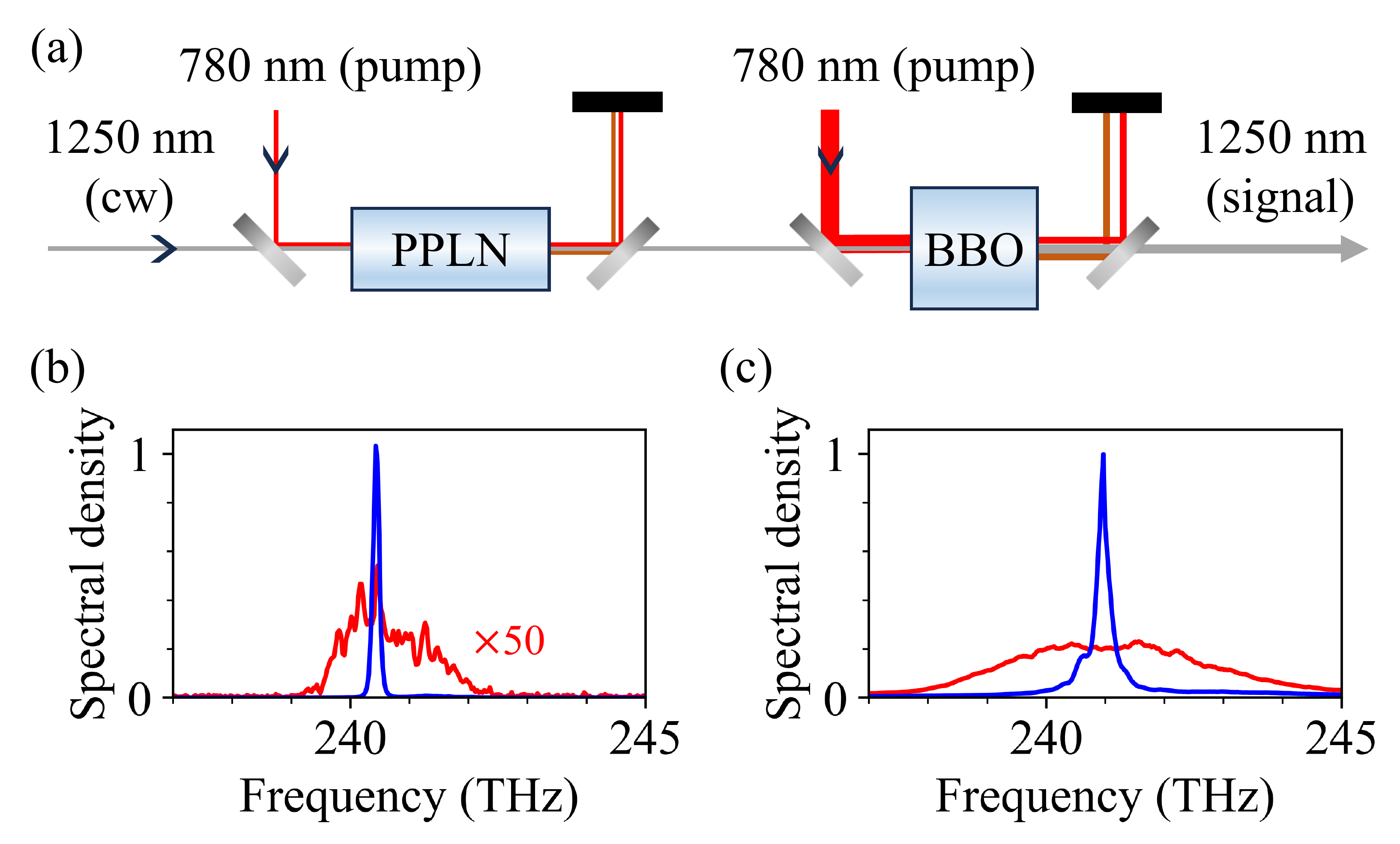}
\caption{(a) Schematic of the double-stage OPA, with a PPLN then a BBO crystal. The 780~nm pump amplifies 1250~nm (grey) and 2080~nm (brown). A cw laser at 1250~nm is used for injection seeding of the OPA. (b) Spectra, normalized to the peak, of 1250~nm pulse after PPLN with/without seeding (blue/red lines). (c) Same, after the BBO crystal.}
\label{fig:image2}
\end{figure}

\begin{figure}[ht]
\centering
\includegraphics[width=\linewidth]{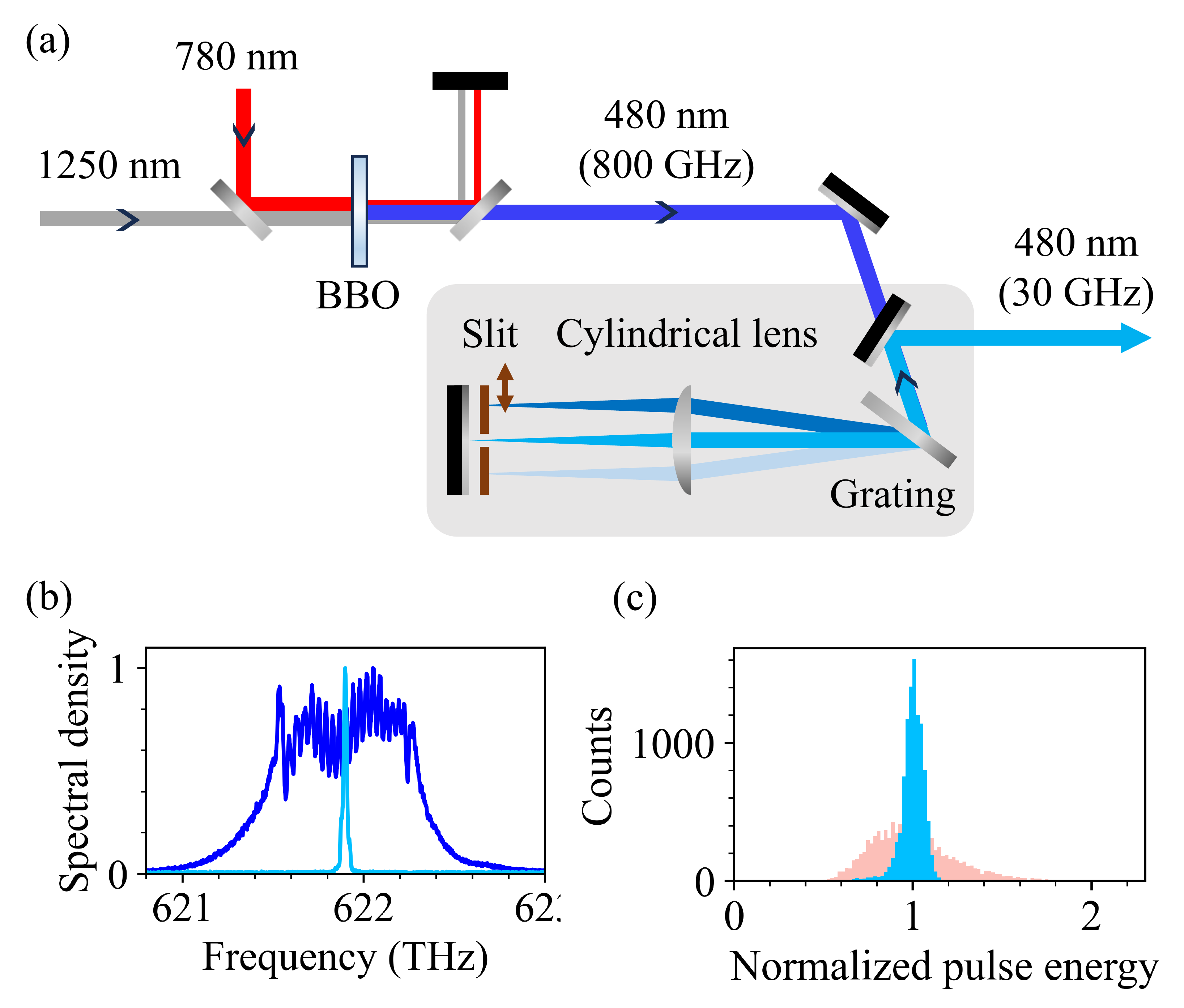}
\caption{(a) Schematics of the SFG and spectral shaper. The 1250~nm and 780~nm pulses are frequency-summed by the BBO crystal to generate 480~nm pulses with a bandwidth of 800 GHz. The spectral shaper (gray box) reduces the bandwidth down to 30 GHz. (b) Spectra of the 480~nm pulse before/after spectral shaping (dark/light blue). The apparent fringes on the dark blue spectrum are a measurement artifact. (c) Histogram of the output pulse energy, with/without injection-seeding (blue/red).}
\label{fig:image3}
\end{figure}

We now present the overall picture of the frequency conversion system shown in Fig.~\ref{fig:image1}(b). 
A commercial device first generates 780~nm pulses for (i) exciting atoms from $5S$ to $5P$, and (ii) pumping the home-built frequency conversion system generating the 480~nm pulses performing the $5P$ to Rydberg excitation. The two pulses are shone one after the other (780, then 480), with a controllable delay.
The frequency conversion system is composed of two building blocks.
Firstly, an optical parametric amplifier (OPA), pumped by the 780~nm (384~THz) pulse, boosts by 90~dB and during 2~ps a cw seed tunable in the range of $1245-1256$~nm ($238.8-241$~THz). This will be detailed in Fig.~\ref{fig:image2}. 
Secondly, a $479.5-481$~nm pulse ($623-625.6$~THz) is generated by the sum-frequency generation (SFG) of $\sim$1250~nm and 780~nm pulses, and will be described in Fig.~\ref{fig:image3},
The tuning range of the OPA and SFG allows to cover the entire Rydberg manifold from $n =35$ to the ionization threshold.
At the output of the SFG, the bandwidth of the 480~nm pulse is 800 GHz, too large to resolve a single Rydberg state (typically separated by 100~GHz). As a solution, the 480~nm pulse (and also optionnally the 780~nm pulse) finally passes through spectral shaping modules to cut the bandwidth down to 30~GHz. This completes our general overview of the laser system. 

Next, we will discuss about our pulsed laser source.  It is composed of an actively mode-locked Titanium:sapphire (TiSa) oscillator (SpectraPhysics, Mai Tai), followed by a chirped-pulse amplification unit (SpectraPhysics, Spitfire Ace) delivering 780~nm pulses at a repetition rate of 1~kHz with an energy of 5~mJ. The energy is very stable from pulse to pulse, with a fluctuation of 0.5~\% (rms) measured with a pyro-electric sensor. By design of the CPA unit, the pulse spectrum is quasi-rectangular, with a width of 800~GHz. We have measured its temporal duration by auto-correlation and obtained a width of 2~ps (FWHM). We note that this is larger than expected from a Fourier-transform limited pulse (1.1~ps, assuming a time-bandwidth product of 0.89 for a rectangular spectrum). We attribute this to a 0.5~ps$^2$ chirp originating from an imperfect alignment of the compressor at the end of the CPA unit. As discussed later, it will have no sizeable influence in the final 480~nm pulse.  
The high peak power (2.5~GW) provided from this laser source will efficiently drive the follow-up single-pass, free-space, non-linear conversion system (and also easily damage many optical components if focused too much).

We now move to the OPA block of Fig.~\ref{fig:image1}(b), amplifying a 1250~nm seed with the 780~nm pump described in the previous paragraph. As shown in Fig.~\ref{fig:image2}(a), we use two stages based on, first, a periodically-poled lithium niobate (PPLN) crystal (effective nonlinear coefficient $d_{\rm eff}= 14$~pm/V, measured damage threshold at 500~MW/cm$^2$), followed by a $\beta$-Barium Borate (BBO) crystal ($d_{\rm eff}= 2$~pm/V, damage at 4~TW/cm$^2$). The motivation for this architecture is that the PPLN, with its high gain but lower output, acts as a first-stage amplifier, while the BBO that can be pumped harder allows to further boost the 1250~nm peak power. We now give more details to each of the amplifiers. For the PPLN, it is designed with a poling period of 20~$\mu$m providing quasi-phase matching of the 780~nm pump with the 1250~nm \textit{signal} and 2080~nm \textit{idler}, which can be precisely tuned by temperature to cover all required wavelengths. The crystal length was chosen to be as long as 40~mm, to provide higher amplification. Due to dispersion, this limits the phase-matching bandwidth to 2.5~THz, but more importantly to a temporal walk-off of 20~ps between the signal and the pump. To ensure a reasonable temporal overlap in the PPLN, we stretched the pump to 9~ps by spectral shaping. We choose a pump energy of 5~$\mu$J, giving a peak intensity of 127~MW/cm$^2$, below the damage threshold of the PPLN surface. The second BBO-based amplifier, taking over the PPLN to further boost the 1250~nm pulse, is designed as follows. The crystal is cut for type-I phase-matching, with a length of 10~mm. It gives a temporal and spatial walk-offs of 0.2~ps and 0.5~mm, with negligible impact given the pump duration of 2~ps and the beam diameter of 2~mm. The pump energy is 1.6~mJ, with a peak intensity of 50~GW/cm$^2$. We observed that, for increasing 1250~nm input, the pump could be strongly depleted, and even observed back-conversion of the 1250 to 780, indicating that the frequency conversion can even get \textit{too} efficient. We simply adjusted the input 1250~nm to remain in a good regime. This concludes our description of the amplifiers design.

First, we operated the OPA system only pumped by the 780~nm pulse, without injection seeding of the PPLN.
In this case, the PPLN amplifies the always existing quantum fluctuation of the 1250~nm field, a regime called optical parametric generation (OPG).
In fact, we used this regime in our previous study using a commercial frequency-conversion device (TOPAS-C, Light Conversion)~\cite{chew2022ultrafast}. The red lines in Fig.~\ref{fig:image2}(b,c) are the spectra of the 1250~nm signal pulse after the PPLN and the BBO stages, with a FWHM bandwidth of 2.5~THz and 4~THz, respectively. This is in perfect agreement with both amplifiers phase-matching bandwidth, but clearly the output pulse is not Fourier-limited with respect to a 2~ps duration. In addition, pulse-to-pulse energy fluctuations at the output of the amplifiers are measured to be 4.5~\%, significantly larger than the pump fluctuations of 0.5~\%. Both observations are attributed to the amplifiers being seeded by quantum fluctuations, which can be solved by actively seeding the OPA units with a laser. 

We then describe the OPA system performance when seeded by a 1250~nm cw laser. The latter has an average power of 50~mW, corresponding to an energy of 500~fJ (3 million photons) in the 10~ps window of the PPLN pump. The seed is injected collinearly with the 780~nm pump in the PPLN and amplified up to 100~nJ (53~dB gain). The bandwidth of the output decreases from 2.5~THz to 100~GHz with the seeding (blue spectrum in Fig.~\ref{fig:image2}(b)), now consistent with the pump bandwidth. Next, the BBO further amplifies the pulse to an energy of 300~{\textmu}J (35 dB gain), with a bandwidth of 300 GHz (blue spectrum in Fig.~\ref{fig:image2}(c)), much narrower than the 4 THz without seeding. The bandwidth of the output 1250 is even narrower than the 800~GHz pump bandwidth. We explain this from the above-mentioned chirp of the pump, increasing the pump duration to 2~ps, which then corresponds to a $200-400$~GHz Fourier-limited bandwidth (depending on the precise temporal profile), in good agreement with our observation. We note that, being seeded by the cw-laser, the output 1250~nm is not chirped (the unused 2080~nm output is chirped instead).

The next step is the sum-frequency generation of 780 and 1250 to generate the final 480~nm pulse, see Fig. \ref{fig:image3}(a). We use again a BBO crystal, with a 1~mm thickness giving negligible temporal and spatial walk-off. 
From the 300~$\mu$J of 1250~nm and 800~$\mu$J of 780~nm, we generate 200~$\mu$J, corresponding to a 25~\% conversion efficiency in terms of 1250~nm photon numbers, which could be improved in future works with a thicker BBO. The output bandwidth, shown in Fig.~\ref{fig:image3}(b), was 800~GHz and determined by the pump bandwidth (more precisely, the convolution of the 1250 and 780 spectra), which is too broad to resolve the individual Rydberg states spaced by $\sim$100 GHz.

A spectral shaping system is used to reduce the bandwidth of the pulse from 800 GHz (Fig.~\ref{fig:image3}(c), dark blue spectrum) to 30 GHz (light blue). As shown in Fig.~\ref {fig:image3}(a), the pulse passes through a grating (3600 grooves/mm) which spatially disperses the frequency components, with a diffraction efficiency of 90$\%$. We select a narrow frequency band using a slit, with a resolution of 30 GHz limited by the diffraction limit (set by the beam size on the grating and its resolving power). This is now sufficient to resolve the Rydberg states. However, in our previous study, we observed large pulse-to-pulse energy fluctuation of 30$\%$ after this spectral shaping~\cite{chew2022ultrafast}. Having now seeded the OPA systems, these fluctuations are successfully suppressed down to 6.2~\%, as shown in the histogram of Fig.~\ref{fig:image3}(c), to be compared to 30~\% in the absence of seeding. The remaining fluctuations are decomposed as follows: we observe 2.3~\% after the OPA, then 3.1~\% after the SFG and finally gradually increasing to 6~\% when narrowing the bandwidth with the spectral shaping system. This concludes our description of the improved frequency-conversion system now generating a much more stable 480~nm pulse.

\begin{figure}[ht]
\centering
\includegraphics[width=\linewidth]{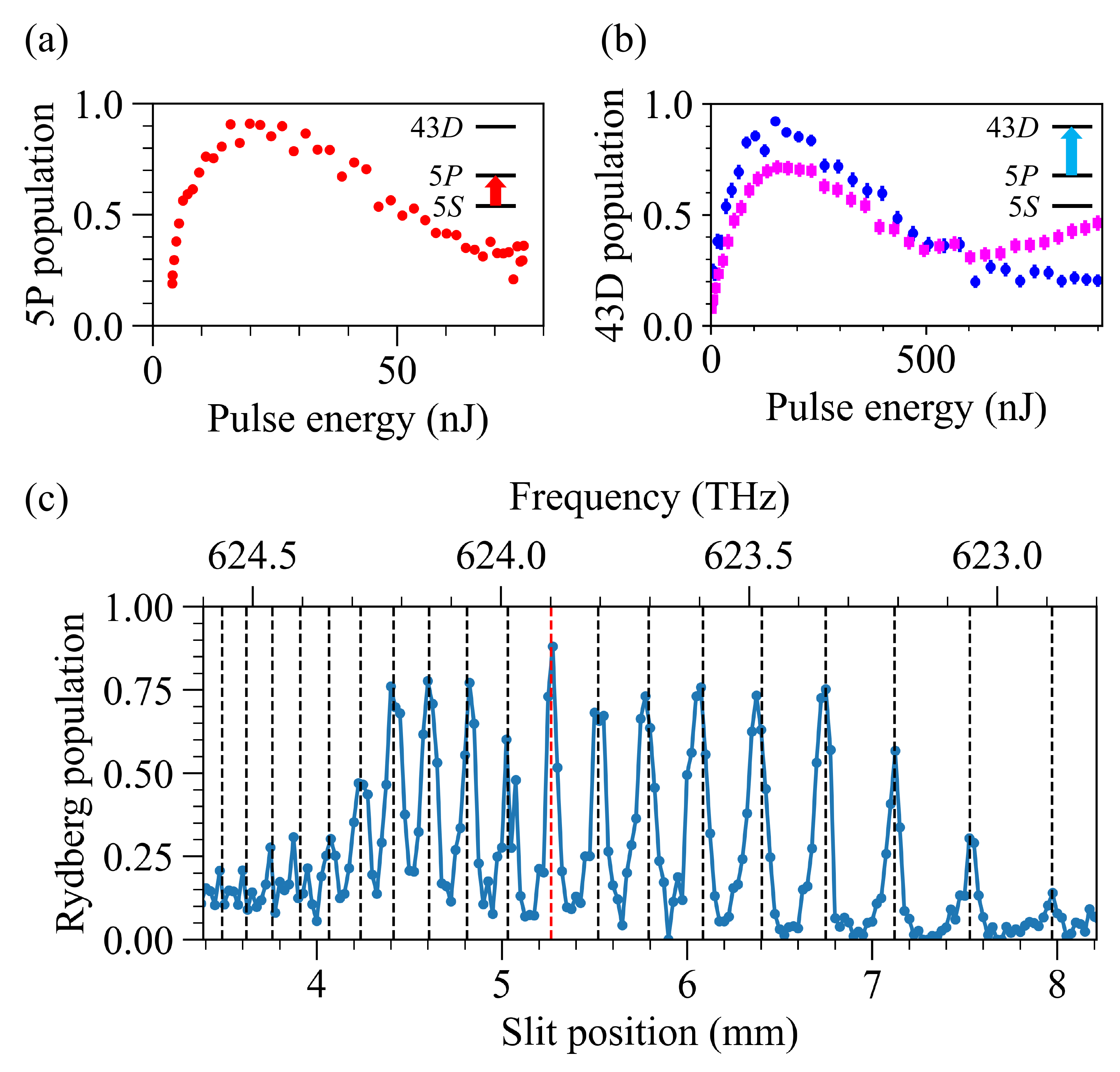}
\caption{(a) Ultrafast Rabi oscillation between the $5S$ and $5P$ states, as a function of the 780~nm pulse energy. (b) Same, between the $5P$ and Rydberg $43D$ states, with the 480~nm pulse. Magenta: previous setup without injection-seeding, blue: upgraded system with injection-seeding. (c) Spectrum of the Rydberg $nD$ series, obtained by scanning the position of the slit in the spectral shaper (1~mm = 350 GHz). Vertical lines denote the expected Rydberg resonances (red is $43D$).}
\label{fig:image4}
\end{figure}

We now demonstrate the efficient ultrafast excitation of atoms to Rydberg states with this improved pulsed laser system. 
We use a cold-atom experimental system, described previously~\cite{chew2022ultrafast}, providing $^{87}$Rb atoms trapped in optical tweezers. A single atom was used for the data shown below. 
The atom is prepared in the spin-polarised $5S$ state $\ket{F=2,m_{F}=2} = \ket{m_{L}=0; m_{S}=1/2; m_{I}=3/2}$ by optical pumping. Two laser pulses are sequentially shone for excitation first to the $5P$ and then to a Rydberg $nD$ state. Polarization of both pulses are $\sigma^{+}$ to always remain in the orbital \textit{stretched} state $m_L = L$, so we drop the notations of spin and orbital angular projection afterwards. 
The first 780 nm pulse has a large beam diameter of 1 mm, to minimize alignment sensitivity since available power is not a limit here. Its duration was set to 10~ps (increased from 2~ps with a spectral shaper), much shorter than the 25~ns lifetime of the $5P$ state allowing to neglect it. 
Figure~\ref{fig:image4}(a) shows the population in the $5P$ state displaying a coherent Rabi oscillation while varying the pulse energy (at constant pulse duration). The $5P$ population is measured by converting it to an atom-loss by subsequently shining the 480~nm pulse at its full bandwidth, photo-ionizing the atom. We observe a Rabi $\pi$-pulse with a maximum population of 90~\%, obtained at a pulse energy of 22~nJ, to be compared with an ab-initio estimate of 10~nJ, a reasonable agreement. The imperfections are attributed partly to state preparation and measurement errors, as well as to polarization error of the 780~nm pulse. The pulse-to-pulse energy fluctuation of 0.5$\%$ is predicted to have negligible contribution.

The next step is bringing the atom from $5P$ to the Rydberg state, by irradiating the 480~nm after a delay of 200~ps (which could be further reduced). With a 30~GHz bandwidth (after spectral shaping), this pulse has a Fourier-limited duration of 15~ps (assuming a Gaussian shape). It is focused down to a size of 20~$\mu$m to maximize the intensity given the limit in available power. 
We first perform spectroscopy of the Rydberg manifold, by scanning the central frequency of the 480~nm pulse with the spectral shaping system described previously. The population in the $nD$ Rydberg state is observed as an atom-loss and shown in Fig.~\ref{fig:image4}(c), together with the calculated energy of the Rydberg state (dashed vertical line). We observe clearly resolved states from $n = 35$ (at 622.9~THz) to $n=51$ (at 624.9~THz), limited by the finite bandwidth of the 480~nm pulse before spectral shaping. Lower states could also be observed by tuning the seed laser and OPA system, while higher states would hardly be resolved. We then selected the $43D$ state (red line) to perform a Rabi oscillation, whose results is shown as blue dots in Fig.~\ref{fig:image4}(b). We observe a Rabi $\pi$-pulse with a population of $\sim$90~\% for an energy of 170~nJ, in agreement with the numerical estimate predicting a $\pi$-pulse with 160~nJ.
In contrast to previous experiment (magenta) performed with the unseeded commercial TOPAS system~\cite{chew2022ultrafast}, we observe a higher excitation probability: 90~\% instead of 75~\%. The improvement is attributed to injection-seeding the OPA system, which strongly reduced the pulse-to-pulse energy fluctuation. The latter now contributes an estimated $(\pi \epsilon/4)^2 \sim 0.2 \, \%$ decrease of population, negligible compared to the error from the imperfect preparation in the $5P$ state ($\sim 10$~\%). We also calculated that nearby Rydberg states play no role at the percent level (also experimentally supported by the well-resolved lines in the spectrum).

Finally, we discuss future directions to further improve ultrafast excitation, beyond the immediate task of improving state preparation. A first idea is to increase the 480~nm energy with a better use of the pump. For example, we expect that the SFG efficiency could be easily doubled with a longer crystal. Even more promisingly, we are pursuing a revision of the spectral shaping system where, instead of \textit{cutting} the spectrum (currently giving a 90~\% decrease of energy), we would rather \textit{compress} it~\cite{courtney2019}. Another development is to switch the pump source from a TiSa-based to a Nd:YAG-based system, delivering higher pump energy. 
A second approach, unlocked by more laser power, is to use more advanced excitation scheme than a Rabi $\pi$-pulse, such as a rapid adiabatic passage with a chirped pulse. 

\begin{backmatter}
\bmsection{Acknowledgements} We thank T. Taira, H. Ishizuki, A. Kausas, B. Bruneteau, F. Cassouret, and Y. Okano for helpful discussions and technical supports. This work was supported by MEXT Quantum Leap Flagship Program JPMXS0118069021, JSPS Grant-in-Aid for Specially Promoted Research Grant No. 16H06289 and JST Moonshot R$\&$D Program Grant Number JPMJMS2269. 
\bmsection{Disclosures} The authors declare no conflicts of interest.
\bmsection{Data availability} Data underlying the results presented in this paper are not publicly available at this time but may be obtained from the authors upon reasonable request.
\end{backmatter}



\end{document}